\documentclass[useAMS,usenatbib]{mn2e}
\pdfoutput=1
\usepackage{aas_macros}
\usepackage{epsfig}
\usepackage{amsmath}
\usepackage{amssymb}
\usepackage{comment}
\usepackage{caption}
\usepackage{leftidx}
\usepackage{natbib}
\usepackage[rgb]{xcolor}
\definecolor{MyGreen}{rgb}{0.0,0.6,0.3}
\definecolor{MyPurple}{rgb}{0.6,0,0.3}
\usepackage{fancyref}
\usepackage{hyperref}
\hypersetup{colorlinks=true,citecolor=MyGreen,linkcolor=MyPurple,urlcolor=blue}

\def\beq{\begin{equation}}
\def\eeq{\end{equation}}
\def\ba{\begin{eqnarray}}
\def\ea{\end{eqnarray}}
\def\bal{\begin{align}}
\def\eal{\end{align}}
\def\bxi{{\mbox{\boldmath $\xi$}}}

\begin{document}

\title[Inverse Tides] {Inverse Tides in Pulsating Binary Stars}

\author[J. Fuller]{
Jim Fuller$^{1}$\thanks{Email: jfuller@caltech.edu}
\\$^1$TAPIR, Mailcode 350-17, California Institute of Technology, Pasadena, CA 91125, USA
}

\label{firstpage}
\maketitle

\begin{abstract}

In close binary stars, the tidal excitation of pulsations typically dissipates energy, causing the system to evolve towards a circular orbit with aligned and synchronized stellar spins. However, for stars with self-excited pulsations, we demonstrate that tidal interaction with unstable pulsation modes can transfer energy in the opposite direction, forcing the spins of the stars \textit{away} from synchronicity, and potentially pumping the eccentricity and spin-orbit misalignment angle. This ``inverse" tidal process only occurs when the tidally forced mode amplitude is comparable to the mode's saturation amplitude, and it is thus most likely to occur in main sequence gravity mode pulsators with orbital periods of a few days. We examine the long-term evolution of inverse tidal action, finding the stellar rotation rate can potentially be driven to a very large or very small value, while maintaining a large spin-orbit misalignment angle. Several recent asteroseismic analyses of pulsating stars in close binaries have revealed extremely slow core rotation periods, which we attribute to the action of inverse tides.
%We conclude with a discussion of the limitations and predictions of the inverse tide theory.

\end{abstract}


\section{Introduction}

Tidal forces between binary stars periodically change the shapes of the stars, and frictional processes within the stars can damp those motions, moving the system to a lower energy configuration. The lowest energy configuration is a circular orbit in which the spins of the stars are synchronized and aligned with the orbital axis \citep{darwin:80,adams:20}. It has long been noted that stars in short-period orbits tend to have circular orbits \citep{meibom:06,torres:10,Shporer2016,vaneylen:16,lurie:17}, and their spins tend to be synchronized. A similar process happens in planets and planetary moons. 

There is an enormous literature on the sources of tidal friction, i.e., the physical processes that convert tidal motion into heat within the star or planet (see \citealt{Ogilvie2014} or \citealt{Barker2020} for recent reviews). These include an effective viscosity produced by convection \citep{zahn:77,goldreich:77,Ogilvie2012,Vick2020}, the tidal excitation of traveling gravity waves or inertial waves which are damped by thermal diffusion or non-linear wave breaking \citep{zahn:75,goodman:98,Papaloizou2010,Ogilvie2013,Auclair2015,mathis:15,Guenel2016} and resonantly excited stellar oscillation modes that damp by similar processes (e.g., \citealt{Witte1999,Weinberg2012,burkart:12,Ivanov2013,fuller:17}). In the latter case, it is well known that the energy dissipation rate is proportional to the mode damping rate. 

But what happens when the tidally excited mode is not damped, but rather a self-excited (overstable) oscillation mode whose damping rate is negative? The naive conclusion is that the energy dissipation rate within the star is negative, i.e., the star \textit{loses} energy and the orbit is pumped to a \textit{higher} energy state. The source of this energy is the thermal energy of the star, which is typically channeled into the oscillation mode via a heat engine excitation process such as the $\kappa$-mechanism. In this case, the tidal bulge produced by the mode would not lag behind the motion of the companion, but would rather \textit{lead} in front of it, producing a torque opposite to the usual case of a damped mode.

In this paper, we show that this ``inverse" tidal process can indeed occur, but only under the right circumstances. A proper treatment of the energy flow must include both tidal and linear driving of the stellar oscillation mode, in addition to some form of non-linear damping that saturates the mode energy at a finite level. With these considerations, we show that inverse tidal energy transfer only occurs when the tidally driven pulsation amplitude is similar to the non-linear saturation amplitude. While this limits which pulsating binaries can undergo inverse tidal evolution, recent observations have revealed several systems that match the observational predictions we lay out below, providing a compelling case for the operation of this mechanism in Nature.

\section{Mode-Orbit Interaction}
\label{modeorbit}

Just as tidal forces can excite stellar oscillation modes, a star's self-excited oscillations can gravitationally interact with the orbit of a companion star. We compute this interaction by including tidal forcing, self-excitation, and non-linear damping of the mode's amplitude. In what follows, we consider quadrupole ($\ell=2$) modes in the primary star of mass $M_1$ and radius $R_1$, interacting with a companion of mass $M_2$ in a circular orbit of period $P$ and orbital separation $D$. 

Following the notation of \cite{fullerheartbeat:17}, a stellar oscillation mode of complex amplitude $a$ and angular frequency $\omega$ (measured in the star's rotating frame) evolves according to (see also \citealt{schenk2002})
\beq
\label{adot}
\dot{a} + i \omega a - \gamma a = \frac{i}{2 \omega} \langle \bxi^* \vert - \nabla U \rangle \, .
\eeq
Here, $\gamma$ is the mode growth rate, and the term on the right-hand side of equation \ref{adot} is the tidal driving force, given by 
\beq
\label{tideforce}
\langle \bxi^* \vert - \nabla U \rangle =  \epsilon \frac{G M_1}{R_1^3} \mathcal{D}_{mm'} Q e^{- i \omega_{\rm f} t} \, ,
\eeq
where $\epsilon = (M_2/M_1) (R_1/D)^3$, and the quadrupole moment 
\begin{align}
\label{Q}
Q &= \frac{1}{M_1 R_1^2} \int dV \delta \rho r^2 Y_{\ell m'}^* \nonumber \\
&= - \frac{ (2 \ell+1) R_1}{4 \pi G M_1} \delta \Phi (R_1) \, .
\end{align}
Here, $\delta \rho$ is the normalized Eulerian density perturbation created by the mode, $\delta \Phi(R)$ is its surface gravitational potential perturbation, and $Y_{\ell m'}$ is the spherical harmonic associated with the ($\ell$,$m'$) component of the tidal potential. The factor $\mathcal{D}_{mm'}$ accounts for spin-orbit misalignment and is defined in \cite{lai:12} and in equations \ref{d22}-\ref{d10} in Section \ref{evolution}. The tidal forcing frequency in the rotating frame of the star is
\beq
\label{omf}
\omega_{\rm f} = m' \Omega - m \Omega_s \, ,
\eeq
where $m$ is the azimuthal number of the oscillation mode, $\Omega$ is the orbital frequency, and $\Omega_s$ is the star's spin frequency. 

In the absence of the tidal force, equation \ref{adot} has only a homogeneous solution $a \propto e^{- i \omega t + \gamma t}$ which is simply a freely oscillating mode of frequency $\omega$ and growth rate $\gamma$. Self-excited modes ($\gamma > 0$) grow exponentially until non-linear effects saturate their amplitude. We parameterize this non-linear saturation with an additional non-linear damping
\beq
\label{gammanl}
\gamma_{\rm nl} = -\frac{|a|^2}{a_{\rm nl}^2} | \gamma | \, ,
\eeq
where the non-linear saturation amplitude $a_{nl}$ is a positive real number. This type of linearly driven, non-linearly damped harmonic oscillator is known as a van der Pol oscillator \citep{vanderpol:20,sprott:17}, whose behavior has been well studied. We consider the case with $\gamma \ll \omega$ relevant to stellar oscillations, in which a linearly excited freely oscillating mode reaches an equilibrium amplitude 
\beq
\label{afree}
a = a_{\rm nl} e^{- i \omega t} \, .
\eeq

Including the tidal force, there is also a forced solution to equation \ref{adot}
\beq
\label{aforced}
a = a_{\rm f} e^{- i \omega_{\rm f} t} \, , 
\eeq
where
\beq
\label{af}
a_{\rm f} = a_{\rm tide} \frac{\omega}{\omega - \omega_{\rm f} + i \gamma} \, ,
\eeq
and 
\beq
\label{atide}
a_{\rm tide} = \frac{1}{2} \, \epsilon \, Q \, \mathcal{D}_{m m'} \, \bar{\omega}^{-2}, 
\eeq
and $\bar{\omega}^2 = \omega^2/(GM_1/R_1^3)$.

A mode may simultaneously exhibit both its free (equation \ref{afree}) and forced (equation \ref{aforced}) solution and hence it may simultaneously oscillate at both its natural frequency $\omega$ and a tidal forcing frequency $\omega_{\rm f}$. Usually the free solution is neglected for tidal interactions, because its time-averaged effect will be zero since it does not oscillate at a tidal forcing frequency. However, when including the non-linear damping term of equation \ref{gammanl}, the free and forced solutions are no longer independent, since they interact through the non-linear damping term that depends on the total mode amplitude $|a|^2$. 

Hence, we must simultaneously solve for the forced and free solutions to equation \ref{adot} including a non-linear damping term, 
\beq
\label{adot2}
\dot{a} + i \omega a - \gamma a + \frac{|a|^2}{a_{\rm nl}^2} |\gamma | a = i \omega a_{\rm tide} e^{-i \omega_{\rm f} t} \, .
\eeq
Based upon the solutions above, we assume a solution of the form 
\beq
\label{asol}
a = a_{\rm f} e^{-i(\omega_{\rm f} t + \delta_{\rm f})} + a_{\rm free} e^{-i \omega t} \, ,
\eeq
where $a_{\rm f}$ and $a_{\rm free}$ are real numbers, and $\delta_{\rm f}$ is a phase offset between the tidal forcing and the forced amplitude. Plugging equation \ref{asol} into equation \ref{adot2} and then matching terms with the same time dependence, we obtain
\beq
\label{a1}
(\omega_{\rm f} - \omega - i \gamma) a_{\rm f} + \frac{i |\gamma |}{a_{\rm nl}^2} \big( a_{\rm f}^2 + 2 a_{\rm free}^2 \big) a_{\rm f} = - \omega a_{\rm tide} e^{i \delta_{\rm f}} \, ,
\eeq
and
\beq
\label{a1b}
-i \gamma a_{\rm free} + \frac{i|\gamma|}{a_{\rm nl}^2} (2 a_{\rm f}^2 + a_{\rm free}^2) a_{\rm free} = 0 \, .
\eeq

When substituting equation \ref{asol} into equation \ref{adot2}, there appear additional unbalanced oscillatory terms of the form $i |\gamma| (a_{\rm f}/a_{\rm nl})^2 a_{\rm free} \exp[i( - 2 \omega_{\rm f} t + \delta_{\rm f} + \omega t)]$ and $i |\gamma| (a_{\rm free}/a_{\rm nl})^2 a_{\rm f} \exp[i( - 2 \omega t + \delta_{\rm f} + \omega_{\rm f} t)]$
. These terms indicate the ansatz of equation \ref{asol} is not perfect, and additional correction terms are required for a complete solution. However, in the limit $|\gamma| \ll |\omega - \omega_{\rm f}|$ we assume below, these terms are small and will not greatly affect the values of $a_{\rm free}$, $a_{\rm f}$, or $\delta_{\rm f}$, which we will verify with numerical experiments in Section \ref{numeric}.

Equation \ref{a1b} always has the trivial solution $a_{\rm free}=0$. The other solution is
\beq
\label{a2}
a_{\rm free}^2 = \frac{\gamma}{|\gamma|} a_{\rm nl}^2 - 2 a_{\rm f}^2 \, .
\eeq
Equation \ref{a2} immediately shows that if the forced amplitude is large compared to the non-linear saturation amplitude, $a_{\rm f}^2 > a_{\rm nl}^2/2$, then the freely oscillating component of the mode ceases to exist, since we have defined $a_{\rm free}^2 \geq 0$. When this occurs, the only solution is $a_{\rm free}=0$. The free solution also does not exist for damped modes ($\gamma < 0$). When the free oscillation does exist, plugging equation \ref{a2} into equation \ref{a1} yields a cubic equation for the forced amplitude,
\beq
\label{a3}
(\omega_{\rm f} - \omega + i \gamma ) a_{\rm f} - 3 i | \gamma | \frac{a_{\rm f}^2}{a_{\rm nl}^2} a_{\rm f} \simeq -\omega a_{\rm tide} (1 + i \delta_{\rm f}) \, ,
\eeq
where we have assumed the phase shift $\delta_{\rm f}$ is small. This is the case for realistic stars, whose growth/damping rates are much smaller than the mode frequencies. Since equation \ref{a3} only applies when $a_{\rm f} \lesssim a_{\rm nl}$ the imaginary components of equation \ref{a3} are small, and so $\delta_{\rm f}$ is small.\footnote{An exception is very close to a mode resonance, $|\omega - \omega_{\rm f}| \lesssim \gamma$, but such fine tuned resonances are highly unlikely and we ignore this case here.}
In this limit, the solution for the forced amplitude and phase shift is
\beq
\label{af2}
a_{\rm f} \simeq a_{\rm tide}  \frac{\omega}{ \omega - \omega_{\rm f}} \, ,
\eeq
and
\beq
\label{df1}
\delta_{\rm f} = \frac{\gamma}{\omega - \omega_{\rm f} } \bigg( \frac{3 a_{\rm f}^2}{a_{\rm nl}^2} \frac{|\gamma|}{\gamma} - 1 \bigg) \, ,
\eeq
and the mode frequency $\omega$ is defined to be positive.

When no free solution exists, we need only solve for a forced solution. In this case, instead of equation \ref{a3}, we have 
\beq
\label{a4}
(\omega_{\rm f} - \omega - i \gamma) a_{\rm f} + i \frac{| \gamma |}{\gamma} \frac{a_{\rm f}^2}{a_{\rm nl}^2} a_{\rm f} = -\omega a_{\rm tide} (1 + i \delta_{\rm f}) \, ,
\eeq
where we have again assumed that the imaginary terms are small due to small mode growth rates. Then we again recover equation \ref{af2} for the forced amplitude, and the phase shift is
\beq
\label{df2}
\delta_{\rm f} = \frac{\gamma}{ \omega - \omega_{\rm f} } \bigg( 1- \frac{|\gamma|}{\gamma} \frac{ a_{\rm f}^2}{a_{\rm nl}^2} \bigg) \, .
\eeq

%To summarize, we define the degree of forced non-linearity,
%such that the phase shift is
%\begin{align}
%\label{df}
%\delta_{\rm f} &= \frac{\gamma}{\omega-\omega_{\rm f}} \bigg( \frac{3}{2} \frac{|\gamma|}{\gamma} x_{\rm f}^2 - 1 \bigg) \quad , \quad x^2 < 1 \nonumber \\
%&= \frac{\gamma}{\omega-\omega_{\rm f}} \bigg( 1 - \frac{1}{2} \frac{|\gamma|}{\gamma} x_{\rm f}^2 \bigg) \quad , \quad x^2 > 1 \, .
%\end{align}

The sign of $\delta_{\rm f}$ is crucial because it determines the phase lag of the tidally forced solution, and hence determines the sign of the torque upon the orbit. The orbital energy change rate is
\beq
\dot{E}_{\rm orb} = - \int dV \delta \rho \frac{\partial U}{\partial t} \, .
\eeq
Some algebra demonstrates this can be written
\begin{align}
\label{edot}
\dot{E}_{\rm orb} &= -2 \epsilon \frac{GM_1^2}{R_1} \omega_{\rm f} Q \mathcal{D}_{m m'} a_i \nonumber \\
&= 2 \epsilon \frac{GM_1^2}{R_1} \omega_{\rm f} Q \mathcal{D}_{m m'} a_{\rm f} \sin \delta_{\rm f} \nonumber \\
&\simeq \epsilon^2 Q^2 \mathcal{D}_{m m'}^2 \bar{\omega}^{-2}\frac{GM_1^2}{R_1} \frac{\omega_{\rm f}\omega}{(\omega_f-\omega)^2} \gamma_{\rm f}
 \, ,
\end{align} 
with
\begin{align}
\label{gammaef}
\gamma_{\rm f} &= \left( \frac{3}{2} x_{\rm f}^2 - 1 \right) \gamma \, , \quad  \quad \gamma > 0 \, \, {\rm and} \, \,  x_{\rm f}^2 < 1 \nonumber \\
&= \left( 1 - \frac{1}{2}\frac{|\gamma|}{\gamma} x_{\rm f}^2 \right) \gamma \, ,  \quad \quad x_{\rm f}^2 > 1 \, \, {\rm or} \, \,  \gamma < 0 \, ,
\end{align}
and we have defined the forced amplitude non-linearity
\beq
\label{xf}
x_{\rm f} = \sqrt{2}\frac{a_{\rm f}}{a_{\rm nl}} \, .
\eeq
All the terms in the last line of equation \ref{edot} are positive (including $\omega$ which is positive by definition), except for $\omega_{\rm f}$ and $\gamma_{\rm f}$, so the sign of those two terms determines the direction of energy transfer. 

For simplicity, let us consider an aligned prograde orbit and a non-rotating star such that $\omega_{\rm f} = 2 \Omega$ and is positive. For damped modes, we use the second line of equation \ref{gammaef} to see that $\gamma_{\rm f} < 0$, so we find orbital decay ($\dot{E}_{\rm orb} < 0$) as orbital energy is used to spin up the star, as expected. For unstable modes, consider first the case $x_{\rm f}^2 \ll 1$, i.e., very small forced mode amplitudes. In this case, $\gamma_{\rm f} \simeq -\gamma$ is negative since $\gamma > 0$ for a driven mode. Then $\dot{E}_{\rm orb} < 0$ and the orbit decays to a lower energy state, which is the same direction as the usual orbital evolution. For $x_{\rm f}^2 \gg 1$, we also obtain a negative value of $\gamma_{\rm f}$, and again the orbital evolution proceeds in the usual direction, though now at a faster rate due to the stronger non-linear damping.

However, for $x_{\rm f} \sim 1$, we see from equation \ref{gammaef} that the value of $\gamma_{\rm f}$ becomes positive. In this case, \textit{energy is transferred to the orbit from the star}. This only occurs in the range $1 > a_{\rm f}^2/a_{\rm nl}^2 > 1/3$, i.e., when the forced mode amplitude is slightly less than the non-linear saturation amplitude. In this situation the flow of energy and AM is opposite to the usual case, a process we refer to as ``inverse tides." The source of energy is ultimately the star's thermal energy, which is channeled into oscillations via a linear instability, e.g., the $\kappa$-mechanism that operates in SPB stars \citep{dziembowski:93} or the convective flux blocking mechanism in $\gamma$-Dor stars \citep{guzik:00}. The oscillation energy and angular momentum (AM) is then be transferred to the orbit via gravitational interaction with the companion's tidal gravitational field.

We note that equations \ref{af2}, \ref{df1}, and \ref{df2} are only valid when $\delta_{\rm f}$ is small. This is violated when a mode is extremely close to resonance such that $\gamma \lesssim |\omega_{\rm f} - \omega|$, or when its forced amplitude is very large such that $(a_{\rm f}/a_{\rm nl})^2 \gamma/\omega \gtrsim 1$. However, in general, the orbital energy change can be written 
\beq
\label{edot2}
\dot{E}_{\rm orb} = 4 M_1 R_1^2 \omega_{\rm f} \omega a_{\rm f}^2 \gamma_{\rm f} \, ,
\eeq
with $\gamma_{\rm f}$ still evaluated from equation \ref{gammaef}. While the solution for $a_{\rm f}$ can differ from equation \ref{af2}, $a_{\rm f}^2$ is a positive number, so we see that the orbital evolution is still determined by the sign of $\gamma_{\rm f}$.

In the limit $(\gamma/|\omega-\omega_{\rm f}|) x_{\rm f}^2 \gg 1$, we find that the forced amplitude is $a_{\rm f} \simeq - (\omega a_{\rm tide} a_{\rm nl}^2/|\gamma|)^{1/3}$, and
\beq
\label{edot3}
\dot{E}_{\rm orb} \simeq - 4 M_1 R_1^2 \left(\frac{\omega}{|\gamma|}\right)^{1/3} \big(a_{\rm tide}^4 a_{\rm nl}^2 \big)^{1/3} \omega^2 \omega_{\rm f} \, .
\eeq
In this case, the orbital energy evolves in the usual direction. However, due to the enhanced non-linear damping, the decay rate scales much differently than usual, scaling as $a_{\rm tide}^{4/3}$ instead of $a_{\rm tide}^2$, and the decay rate is insensitive to the mode detuning $\omega_{\rm f} - \omega$. It is not clear if this situation occurs in realistic stellar binaries. Additionally, the non-linear damping model of equation \ref{gammanl} likely breaks down for $a \gg a_{\rm nl}$, changing the scaling of equation \ref{edot3}.

\subsection{Numerical Verification}
\label{numeric}

To verify our analytical results, we perform a numerical integration of the coupled orbital and mode amplitude evolution. From equation \ref{adot2}, the real and imaginary components of the mode amplitude evolve according to
\beq
\label{brdot}
\dot{b}_r = \big( \sigma- m \dot{\Phi} \big) b_i + \gamma \bigg[ 1 - \frac{|\gamma|}{\gamma} \frac{b_r^2 + b_i^2}{a_{\rm nl}^2} \bigg] b_r \, ,
\eeq
\beq
\label{bidot}
\dot{b}_i = - \big( \sigma - m \dot{\Phi} \big) b_r + \gamma \bigg[ 1 - \frac{|\gamma|}{\gamma} \frac{b_r^2 + b_i^2}{a_{\rm nl}^2} \bigg] b_i + \omega a_{\rm tide} \, ,
\eeq
where $b = a e^{i m \Phi}$, $\sigma = \omega + m \Omega_s$, and $\Phi$ is the orbital phase. Following \cite{fullerwd:11}, the orbital distance and phase evolve according to
\beq
\label{ddot}
\ddot{D} = - \frac{G(M_1+M_2)}{D^2} + D \dot{\Phi}^2 - \frac{6 G(M_1+M_2)R^2}{D^4} \mathcal{D}_{m m'} Q b_r \, ,
\eeq
\beq
\label{phidot}
\ddot{\Phi} = - \frac{2 \dot{D} \dot{\Phi}}{D} - \frac{2 m G(M_1+M_2)R^2}{D^5} \mathcal{D}_{m m'} Q b_i \, .
\eeq
In the phase space expansion used for the mode amplitude equations, each physical mode is the sum of a mode (of azimuthal number $m$, frequency $\omega_\alpha$, and eigenfunction $\bxi_\alpha$) and its complex conjugate (with azimuthal number $-m$, frequency $-\omega_\alpha$, and eigenfunction $\bxi_\alpha^*$). The amplitude equations \ref{brdot} and \ref{bidot} are identical for each component of the mode. Both components have been included in equations \ref{ddot} and \ref{phidot}, which increases the last terms in each equation by a factor of two.

\begin{figure}
\includegraphics[scale=0.35]{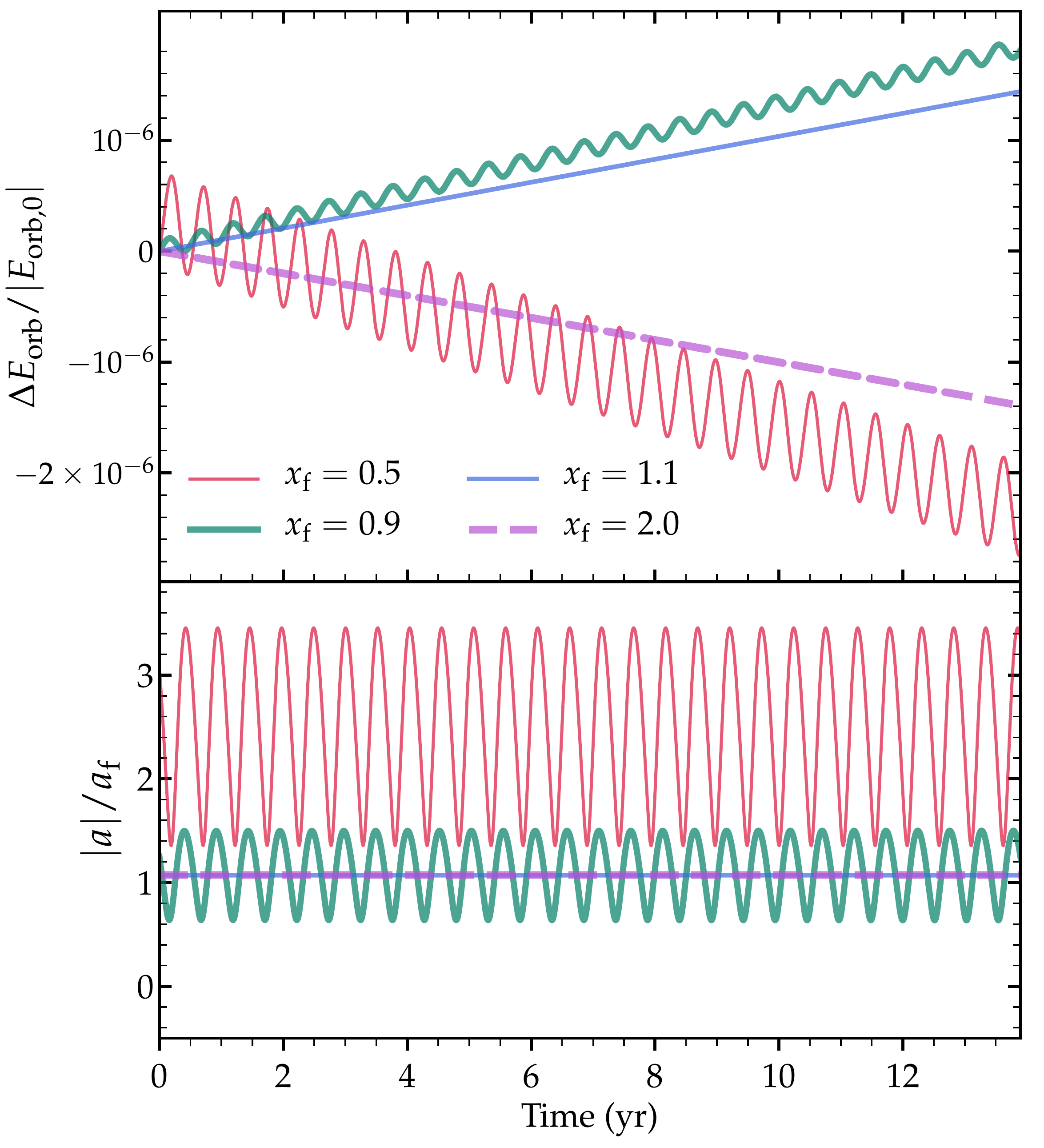}
\caption{\label{OrbDrive} Top: fractional change in orbital energy $\Delta E_{\rm orb}/|E_{\rm orb,0}|$, as a function of time, for a binary star interacting with a self-excited oscillation mode in the primary. Each line corresponds to a different ratio of the forced amplitude to the non-linear saturation amplitude, $x_{\rm f}$ (equation \ref{xf}). Modes with $x_{\rm f} \approx 1$ drive orbital expansion via inverse tides. Bottom: mode amplitude $|a|$ as a function of time, demonstrating the beating between the free and forced component of the mode for low values of $x_{\rm f}$. 
}
\end{figure}

To solve equations \ref{brdot}-\ref{phidot}, we consider a non-rotating $M_1 = 4 \, M_\odot$ primary in a circular orbit with a $M_2 = 2 \, M_\odot$ secondary and orbital period $P = 2.5 \, {\rm days}$. We construct a stellar model with the MESA stellar evolution code \citep{paxton:11,paxton:13,paxton:15,paxton:18,paxton:19} and evolve it about half way through the main sequence, to a radius of $3.4 \, R_\odot$ and surface temperature $T_{\rm eff} = 13,400 {\rm K}$. We then compute non-adiabatic $\ell=2$ oscillation modes with GYRE \citep{townsend:13}, evaluating their frequencies $\omega$, growth rates $\gamma$, and quadrupole moments $Q$. We indeed find unstable g~modes with periods $P \simeq P_{\rm orb}/2 \simeq 1.25 \, {\rm days}$ that are resonant with tidal forcing from the $\ell=m'=2$ component of the tidal potential.
%We consider a non-rotating primary, with a circular aligned orbit with $\mathcal{D}_{m m'} = \mathcal{D}_{22} = 1$. 

We choose the g~mode from the model that is closest to the tidal forcing frequency $\omega_{\rm f} = 2 \Omega_{\rm orb}$, which has  $Q \sim 10^{-5}$. We integrate equations \ref{brdot}-\ref{phidot}, for different values of the non-linear saturation value $a_{\rm nl}$, corresponding to different values of $x_{\rm f}$ from equation \ref{xf}.
%As discussed above, we expect energy transfer from the orbit to the star's spin for small or large values of $x_{\rm f}$, but we expect energy transfer from the star to the orbit for $x_{\rm f} \approx 1$. 
Figure \ref{OrbDrive} shows an example of how the orbital distance and total mode amplitude $|a|$ evolve as a function of time. As expected, we find that the orbital energy increases, and hence the orbital distance increases, for values of $x_{\rm f} \approx 1$. Hence, the orbital AM increases while the star's AM decreases, i.e., the system evolves \textit{away} from spin-orbit synchronicity. For small or large values of $x_{\rm f}$, the orbital distance decreases, moving the system in the usual direction, towards a lower energy state. 

Figure \ref{OrbDrive} also verifies the analytical conclusion that the existence of the free oscillation mode vanishes for $x_{\rm f} > 1$. In this case, the mode amplitude is roughly constant at $|a| \simeq a_{\rm f}$, and the energy transfer rate is steady such that the orbital distance smoothly increases or decreases. However, for $x_{\rm f} < 1$, both the forced and free solutions exist, which oscillate out of phase due to their different frequencies. Hence, the total mode amplitude $|a|$ varies periodically at the beat frequency $|\omega_{\rm f} - \omega|$, as does the orbital distance.

\section{Spin and Orbital Evolution}
\label{evolution}

With an understanding of when inverse tidal processes occur, we now seek to understand the long-term spin and orbital evolution that results. For normal dissipative tides, the system evolves towards the lowest energy state, characterized by a circular orbit, with the spins aligned synchronized with the orbit. For inverse tides, the system evolves towards a \textit{higher} energy state, often causing the system to evolve away from spin-orbit alignment and synchronicity.

One can show that for any tidal process, the orbital eccentricity evolves according to 
\beq
\dot{e^2} = 2(1 - e^2) \bigg[\frac{N}{m} \sqrt{1-e^2} - 1 \bigg] \frac{\dot{L}_{\rm orb}}{L_{\rm orb}} \, ,
\eeq 
where $N \Omega_{\rm orb}$ is the tidal forcing frequency in the inertial frame, and $\dot{L}_{\rm orb}$ is the torque on the orbit. In the limit of small eccentricity, the tidal forcing is strongest at $N=m$, and in this case we find the eccentricity evolves as $\dot{e^2} \simeq -e^2 \dot{L}_{\rm orb}/L_{\rm orb}$. Hence, for an inverse tidal process with positive $\dot{L}_{\rm orb} > 0$, the eccentricity is damped, but eccentricity pumping can occur when $\dot{L}_{\rm orb} < 0$. We shall find both scenarios below, but we will consider only circular orbits, which could be maintained in the case of rapid eccentricity damping due to normal tides in the companion star. For orbits that are already eccentric, further eccentricity pumping may occur for $\dot{L}_{\rm orb} > 0$ due to high frequency inverse tidally excited modes with $N (1- e^2)^{1/2} > m$, or via low frequency inverse tidally excited modes when $\dot{L}_{\rm orb} < 0$. However, in this case one must consider the tidal action due to tidally excited modes at many values of $N$, and we do not investigate this scenario here.

Focusing on circular binaries, we compute the secular evolution of the spin and orbit of a binary system interacting via a stellar oscillation mode. We account for misalignment between the spin and orbital AM vectors, following the procedure and notation of \cite{lai:12}, though we choose a stellar reference frame with opposite $x$ and $y$ directions, such that the star's $z$ axis (defined by its rotation axis) is inclined towards the inertial frame $x$ axis by an angle $\theta_*$. In this frame, the torque along the star's spin axis is
\begin{align}
\label{Tz}
T_{z,*} &= - \frac{3 \pi}{5} T_0 \Omega_s \big ( s_\theta^4 \tau_{20} + s_\theta^2 c_\theta^2 \tau_{10} \big) \nonumber \\
& + \frac{3 \pi}{20} T_0 \Big[ (1 + c_\theta)^4 (\Omega - \Omega_s) \tau_{22} - (1 - c_\theta)^4 (\Omega + \Omega_s) \tau_{-22} \nonumber \\
& + s_\theta^2 (1 + c_\theta)^2 (2 \Omega - \Omega_s) \tau_{12} 
 - s_\theta^2 (1 - c_\theta)^2 (2 \Omega + \Omega_s) \tau_{-12} \Big] \, .
\end{align}
Here, $T_0 = G R_1^5(M_2/D^3)^2$, $c_\theta = \cos \theta$, $s_\theta = \sin \theta$, and $\theta$ is the spin-orbit misalignment angle, which is related to $\theta_*$ by $L \sin \theta = J \sin \theta_*$, where $J$ is the total AM and $L$ is the orbital AM. The time lags $\tau_{m m'}$ define the strength of the torque due to a mode with azimuthal number $m$ excited by the $m'$ component of the tidal potential. Similarly, the torque in the $x$-direction is 
\begin{align}
\label{Tx}
T_{x,*} &= - \frac{3 \pi}{5} T_0 \Omega_s \big ( s_\theta^3 c_\theta \tau_{20} + s_\theta c_\theta^3 \tau_{10} \big) 
 - \frac{3 \pi}{20} T_0 \Big[ 6 s_\theta^3 \Omega \tau_{02} \nonumber \\
& + s_\theta (1 + c_\theta)^3 (\Omega - \Omega_s) \tau_{22} + s_\theta (1 - c_\theta)^3 (\Omega + \Omega_s) \tau_{-22} \nonumber \\
& + s_\theta (1 + c_\theta)^2 (2 - c_\theta) (2 \Omega - \Omega_s) \tau_{12} \nonumber \\
& + s_\theta (1 - c_\theta)^2 (2 + c_\theta) (2 \Omega + \Omega_s) \tau_{-12} \Big] \, .
\end{align}

For a tidally excited oscillation mode, in our notation the corresponding time lag is
\beq
\label{tau}
\tau \simeq - \frac{Q^2}{\bar{\omega}^2} \frac{\gamma_{\rm f}}{(\omega-\omega_{\rm f})^2 + \gamma^2} \, ,
\eeq
Inverse tides occur when the value of $\tau$ is negative, which only occurs when $\gamma_{\rm f}$ is positive. We note here that the overlap of the $m'$ component of the tidal potential with the $m$ spherical harmonic in the star's coordinate frame, $D_{m m'}$ is
\beq
\label{d22}
\mathcal{D}_{\pm 2 2} = \frac{1}{4} (1 \pm \cos \theta)^2 \, ,
\eeq
\beq
\mathcal{D}_{\pm 1 2} = \frac{1}{2} \sin \theta (1 \pm \cos \theta) \, ,
\eeq
\beq
\mathcal{D}_{02} = \mathcal{D}_{20} =  \frac{\sqrt{6}}{4} \sin^2 \theta \, ,
\eeq
\beq
\label{d10}
\mathcal{D}_{10} = -\frac{\sqrt{6}}{2} \sin \theta \cos \theta \, .
\eeq
These factors have already been partially factored out of equations \ref{Tz} and \ref{Tx}, but they still enter through the factor $\gamma_{\rm f}$, because they determine the value of $a_{\rm tide}$ and hence $x_{\rm f}$. The sign of $\tau$ can differ between different $m$ components, depending on their amplitudes, and hence the star may simultaneously exhibit both normal tides and inverse tides.

Using the definition of the spin-orbit misalignment,
\beq
\label{costheta}
\cos \theta = \mathbf{S} \cdot \mathbf{L}/SL \, ,
\eeq
where $\mathbf{S}$ and $\mathbf{L}$ are the spin and orbital AM vectors, the spin-orbit misalignment evolves according to
\beq
\label{thetadot}
\dot{\theta} = \frac{T_{x,*}}{S} + \frac{T_{x,*}}{L} \cos \theta + \frac{T_{z,*}}{L} \sin \theta \, ,
\eeq
which follows from the fact that the total AM in each direction is conserved. In binary stars, the spin AM is typically a couple orders of magnitude smaller than the orbital AM, so the first term on the right hand side of equation \ref{thetadot} often dominates. The small spin AM can allow for rapid changes in the spin-orbit misalignment. The spin AM, measured in the star's frame with $z$-axis along the spin axis, evolves as $\dot{S}_{z,*} = T_{z,*}$ and $\dot{S}_{x,*} = T_{x,*}$.

As discussed above, a circular orbit will usually remain circular even under the actions of inverse tides. In this case, we can simply integrate $S_z$ and $\theta$ in time to determine the coupled spin/orbital evolution, as the orbital distance is determined from AM conservation. However, integrating equation \ref{thetadot} can be problematic because the spin can be driven towards zero such that the first term on the right hand side diverges. So, we instead solve for each component of the star's spin AM measured in an inertial frame, compute the orbital AM from AM conservation, and then compute the spin-orbit misalignment from equation \ref{costheta}. To do this, we use the following relations:
\beq
T_{z,i} = T_{z,*} \cos \theta_* - T_{x,*} \sin \theta_* \, ,
\eeq
\beq
T_{x,i} = T_{x,*} \cos \theta_* + T_{z,*} \sin \theta_* \, ,
\eeq
where $T_{z,i}$ and $T_{x,i}$ are the torques on the star in the inertial frame with total AM in the $z$-direction. In this frame, the total AM is $J = S_z + L_z$, and $S_x + L_x = 0$. The star's spin axis is inclined to the $z$-axis by $\theta_* = \arctan (S_x/S_z)$, and the spin-orbit misalignment angle is 
\beq
\label{tantheta}
\tan \theta = \frac{JS \sin \theta_*}{J S_z - S^2} \, .
\eeq

In some cases, the spin-orbit evolution resulting from inverse tides can be remarkably complex. In what follows, we consider the simple case of tidal evolution arising from a single tidally excited oscillation of azimuthal number $m$, due to forcing by the $m'=2$ component of the tidal potential. This is justified when one resonant oscillation mode dominates the tidal interaction, which could occur over long periods of time if a resonance locking process operates \citep{Witte1999,fullerheartbeat:17}. However, future work should attempt to account for many oscillation modes of different frequencies and various $m$ and $m'$ values, which requires evaluating the appropriate value of equation \ref{tau} for each mode and summing the response. Because the resonance detuning factor will change for each value of $m$ for a rotating star, this could result in very different values of $\tau$ for each $m$ component.

%In what follows, we take the simple approach of setting $Q$, $\omega$, and $\omega_{\rm f}$ equal for all values of $m$, for the $m'=2$ component of the tidal forcing. For simplicity we set $\tau_{10}=\tau_{20}=0$, which is partially justified by the result that the spin will sometimes approach zero, in which case the tidal forcing frequency for the $m'=0$ components approaches zero and interacts with very high-order g modes with tiny values of $Q$. 

\begin{figure}
\includegraphics[scale=0.45]{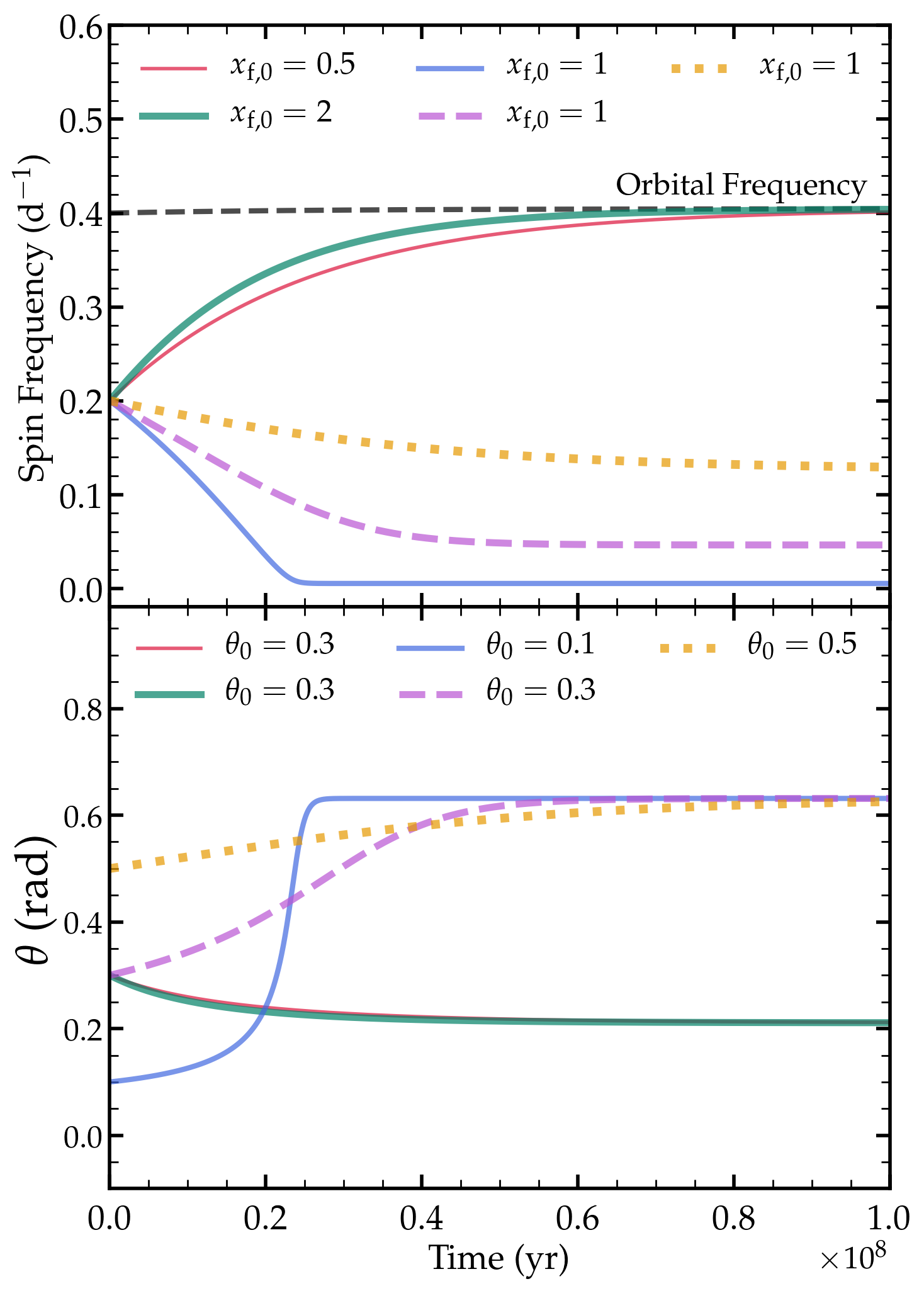}
\caption{\label{OrbEvol} 
Long-term orbital evolution for the binary described in the text, for different values of the non-linear saturation parameter $x_{\rm f,0}$. We consider evolution caused by a single $m\! = \! 2$ oscillation mode driven by the $m' \!= \! 2$ component of the tidal potential. The top panel shows the spin frequency, $\Omega_{\rm s}$, and the bottom panel shows the spin-orbit misalignment $\theta$. For $x_{\rm f,0} \ll 1$ or $x_{\rm f,0} \gg 1$, ``normal" tides cause the system to evolve towards spin-orbit alignment and synchronization. For $x_{\rm f,0} \sim 1$, ``inverse" tides operate, driving the system away from spin-orbit alignment and synchronization. For small initial misalignment angles (blue and purple lines), the star can be driven to extremely small rotation rates, corresponding to spin periods of tens to hundreds of days.
}
\end{figure}

To further simplify our models, we do not account for the changing value of $a_{\rm f}$ and $x_{\rm f}$ due to the changing tidal forcing frequency, and hence the changing mode detuning and forced amplitude via equation \ref{af}. In reality, $x_{\rm f}$ could change greatly, unless the mode amplitude remains nearly constant due to resonance locking. Resonance locking with modes driving inverse tides could potentially operate in real systems, e.g., in stars with sub-synchronous rotation being spun down by inverse tides, such that the tidal forcing frequencies in the star's rotating frame are increasing at the same rate that the gravity mode frequencies naturally increase due to stellar evolution (see discussion in \citealt{fullerreslock:17} and \citealt{fullerheartbeat:17}), but we do not model this explicitly.

However, we do account for the changing spin-orbit misalignment (and hence changing strength of the tidal forcing), by recomputing $\mathcal{D}_{mm'}$ and $\gamma_{\rm f}$ at each time step. For the latter, this is done by setting 
\begin{align}
\label{gammaf2}
\gamma_{\rm f} &= \left( \frac{3}{2} \frac{x_{\rm f,0}^2 \mathcal{D}_{m m'}^2}{{\rm max}(\mathcal{D}_{m m'}^2)} - 1 \right) \gamma \quad , \quad x_{\rm f,0}^2 \mathcal{D}_{m m'}^2 < 1 \nonumber \\
 &= \left( 1 - \frac{1}{2} \frac{x_{\rm f,0}^2 \mathcal{D}_{m m'}^2}{{\rm max}(\mathcal{D}_{m m'}^2)} \right) \gamma \quad , \quad x_{\rm f,0}^2 \mathcal{D}_{m m'}^2 > 1 \, .
\end{align}
Here $x_{\rm f,0} = x_{\rm f} \, {\rm max}(|\mathcal{D}_{m m'}|)/\mathcal{D}_{m m'}$ is a constant representing the forced amplitude without the dependence on misalignment angle. We have defined $x_{\rm f,0}$ this way so that the maximum value of $\gamma_{\rm f}$ occurs at $x_{\rm f,0} =1$ when the value of $|\mathcal{D}_{m m'}|$ is maximized. Thus, a tidally excited mode of azimuthal number $m$ can drive normal or inverse tides, depending on the value of $D_{m m'}$ due to the changing spin-orbit misalignment angle. We leave $x_{\rm f,0}$ as an adjustable parameter to explore different types of orbital evolution. The realistic long-term evolution is likely to be substantially more complex than the results shown below, due to the many tidally excited oscillation modes and their changing values of $x_{\rm f}$. The main goal here is to understand some of the general types of long-term evolution that may be realized, depending on which modes dominate the tidal interaction.

\subsection{Results}
\label{results}

Figure \ref{OrbEvol} shows an example of the long-term coupled spin and orbital evolution of the binary described in Figure \ref{OrbDrive}. In this example, we consider an $m \! = \! 2$ tidally excited oscillation driven by the $m' \! = \! 2$ component of the tidal potential. We begin each orbital integration at $\Omega_s = 0.5 \Omega_{\rm orb}$, for a few different initial spin-orbit misalignment angles $\theta_0$, and  non-linear saturation factors $x_{\rm f,0}$. For small or large values of $x_{\rm f,0}$, inverse tides do not operate, or they only operate over small ranges in $\theta$. In these cases, we find the normal result that the system evolves towards spin-orbit synchronization and alignment. Some misalignment remains in these evolutions because the $m=2$ tidal torque vanishes as $\Omega_s$ approaches $\Omega$, but in a realistic case where normal tides operate via the other terms in equation \ref{Tx} (which do not vanish at $\Omega=\Omega_s$), spin-orbit alignment would eventually be achieved.

However, for values of $x_{\rm f,0} \sim 1$, inverse tides can operate. The action of inverse tides causes the prograde $m \! = \! 2$ modes to transfer AM from the star to the orbit (i.e., $T_{z,*} <0$), opposite to the usual flow of AM, and driving the system away from spin-orbit synchronization to smaller stellar spin frequencies. The inverse tidal torque $T_{x,*}$ is also positive, driving the system away from spin-orbit alignment (equation \ref{thetadot}). As the misalignment angle increases, the factor $\mathcal{D}_{m m'}$ decreases, causing $\gamma_{\rm f}$ to approach zero (equation \ref{gammaf2}). At the critical misalignment angle $\theta_c$ defined by $x_{\rm f}^2 = 2/3$, a stable equilibrium is established. Larger misalignments would cause $\gamma_{\rm f}$ to become negative such that normal tides decrease the misalignment angle back toward $\theta_c$, and smaller misalignments would cause inverse tides to increase the misalignment angle back towards $\theta_c$. The final spin frequency is determined largely by the initial conditions. For small initial misalignment angles (blue line in Figure \ref{OrbEvol}), inverse tides push the system to very small spin rates, where the small spin AM causes a rapid increase in the misalignment angle (first term in equation \ref{thetadot}) until $\theta \simeq \theta_c$ and equilibrium is reached. For larger initial misalignment angles (orange dotted line in Figure \ref{OrbEvol}), the spin frequency is not able to approach zero before $\theta \simeq \theta_c$, and a nearly constant sub-synchronous rotation rate persists indefinitely.

%Like chaotic systems, the evolution is very sensitive to system parameters such as the values of $x_{\rm f,0}$ and $\tau_{m m'}$. Small changes in these parameters not only lead to quantitatively different evolution, but qualitatively different types of evolution. Using $x_{\rm f,0} = 0.9$ in Figure \ref{OrbEvol} leads to a spin frequency of almost exactly zero, with small periodic variations of $\theta$. Experimentation with different relative values of $\tau_{m m'}$, and $x_{\rm f,0}$ yielded a wide range of results, including chaotic and periodic limit cycles, spin-orbit alignment and synchronicity, stable equilbria that remain misalgined and non-synchronized, and runaway of the spin towards very large values. In fact, the end of the evolution for $x_{\rm f,0} = 1.1$ exhibits an escape from the basin of attraction at low spin, and the system then runs away towards very large (and anti-aligned) spin. We find this is another common outcome of these evolutions. In much of the parameter space, the star is either driven towards very small or very large spin rates. 

\begin{figure}
\includegraphics[scale=0.36]{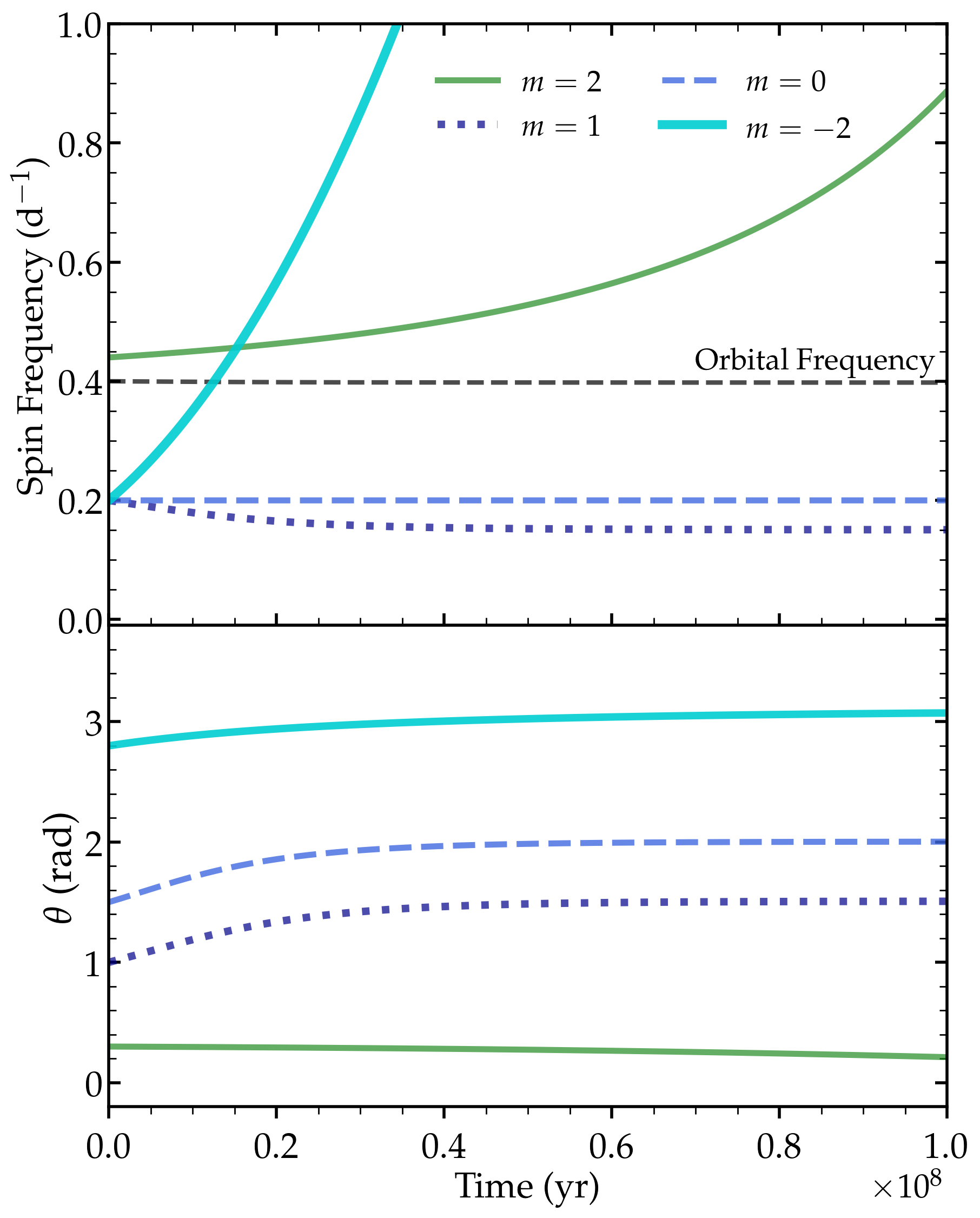}
\caption{\label{OrbEvolDif} 
Evolutionary trajectories of the same binary in Figure \ref{OrbEvol}, but this time for modes with various azimuthal pattern numbers $m$ and initial conditions, and all with with $x_{\rm f,0} = 1$. For aligned super-synchronous initial rotation, we find run-away spin-up via inverse tides with $m \! = \! 2$ modes (green line). For anti-aligned initial rotation, we find run-away spin-up via $m \! = \! -2$ modes. Sub-synchronous initial rotation and inverse tides with $m \! = \! 1$ (purple dotted line) or $m \! = \! 0$ modes (dashed blue line) leads to persistent spin-orbit misalignment and sub-synchronous spin.
}
\end{figure}

Very different outcomes are possible for inverse tides via modes with different values of $m$ or with different initial conditions, shown by a few examples in Figure \ref{OrbEvolDif}. Consider first an $m \! = \! 2$ mode in a star with spin aligned with the orbit, but initially rotating faster than synchronous (green line in Figure \ref{OrbEvolDif}). The resonantly excited mode would have a negative frequency (i.e., a retrograde mode in the star's rotating frame), and the action of inverse tides is to transfer AM from the orbit to the star, spinning up the star to even faster rotation rates. In this case, the value of $T_{x,*}$ is negative so that the system is driven to smaller misalignment angles, and the stable equilibrium at larger misalignment angles where $\theta \simeq \theta_c$ can never be reached. The star thus undergoes run-away spin-up to very large rotation rates. Hence, inverse tidal action can potentially lead to very large or very small stellar rotation rates, depending on the initial conditions.

%Initial conditions also strongly affect the outcome, and Figure \ref{OrbEvolDif} shows four orbital evolutions with $x_{\rm f,0}=1$, but with different initial spin rates or spin-orbit misalignments. In many cases, an initial spin larger than the orbital frequency leads to run-away spin up due to inverse tidal interaction with the $m=2$ mode, as we see for the $\Omega_s/\Omega = 1.3$ curve in Figure \ref{OrbEvolDif}. Physically, this results from inverse tidal interaction with $m=2$ modes that are retrograde in the star's rotating frame (i.e., negative frequency), which transfer AM from the orbit to the star, opposite to the usual flow of AM. Stars with initial values $\Omega_s/\Omega < 1$ typically go through chaotic limit cycles and very small spin frequencies, as discussed above.

Next let us examine the action of inverse tides due to other values of $m$. For significant spin-orbit misalignment ($\theta \! \sim \! 1$) the tidal forcing is strongest for $m \! = \! 1$ modes (purple dotted line in Figure \ref{OrbEvolDif}). Given a sub-synchronous initial rotation, the evolution is quite similar to that examined in Figure \ref{OrbEvol}: inverse tides drive the star to smaller spin rates and higher misalignment angles. Once again, an equilibrium is reached where $\theta \simeq \theta_c$, which occurs at a larger misalignment angle for $m \! = \! 1$ modes. The action of inverse tides driven by $m \! = \! 0$ modes is also similar (dashed blue line in Figure \ref{OrbEvolDif}), except that these modes do not alter the star's spin frequency and only affect the spin-orbit misalignment, which increases towards a larger value of $\theta_c$. Finally, for very large misalignment angles, modes with $m \! = \! -2$ dominate the tidal response (cyan line in Figure \ref{OrbEvolDif}). Inverse tides due to these retrograde modes transfer AM from the orbit to the star, spinning the star up. The value of $T_{x,*}$ is also positive, increasing the spin-orbit misalignment angle towards $\theta \! \approx \! \pi$, so equilibrium where $\theta \! \simeq \! \theta_c$ at smaller misalignments is never reached. The torque from these modes does not vanish when $\Omega_{\rm s} \simeq \Omega$ due to the opposing directions of spin and orbital motion, so the star experiences run-away spin-up in this case as well.

The long-term evolution of realistic star systems undergoing inverse tidal processes could be very complex and should be further examined in future work. Small changes in parameters such as $x_{\rm f}$ and initial conditions lead to qualitatively different types of evolution, so it is not clear what to expect from realistic systems where the value of $x_{\rm f}$ evolves. Moreover, our simplistic orbital evolutions only included tidal evolution due to a single stellar oscillation mode, and hence only included a single term $\tau_{m m'}$ in equations \ref{Tz} and \ref{Tx}. Experimentation with different relative values of $\tau_{m m'}$ yielded a wide range of behaviors, with some evolutions qualitatively similar to Figures \ref{OrbEvol} and \ref{OrbEvolDif}, and others with different behavior such as periodic or apparently chaotic limit cycles. Hence, while normal tides always drive the system towards the lowest energy state (a circular orbit with aligned and synchronous rotation), inverse tides may drive systems to a wide range of dynamic and higher energy states exhibiting eccentricity, obliquity, and asynchronicity.

Finally, we note that the above results may change when accounting for spin precession of the misaligned primary star. For the binary configurations we consider, the spin precession period is orders of magnitude shorter than tidal time scales, and precession will cause the values of $S_x$ and $S_y$ to oscillate, while maintaining constant $\theta$. We have performed a small set of orbital integrations including spin precession of the primary, finding that the spin evolution does not appear to be qualitatively altered, but future work should analyze the effects of spin precession in more detail.

\section{Discussion and Conclusions}

We have demonstrated that an inverse tidal process can occur in binary systems due to tidal forcing of self-excited stellar oscillation modes, but only when the forced mode amplitude is similar to the non-linear saturation amplitude ($x_{\rm f} \sim 1$, see equation \ref{gammaef}). Here we find two common outcomes: evolution towards very large spin frequencies (which can be either aligned or anti-aligned with the orbit), and evolution towards small spin frequencies with persistent spin-orbit misalignment. Whether the former process can actually drive the star towards breakup spin rates is unclear, as it requires interaction with retrograde g~modes (in the star's rotating frame), which couple more weakly with the tidal potential as the spin frequency increases (e.g., \citealt{bildsten:96,lee:97,fullerwd:14}). We are unaware of any very rapidly rotating close binary systems that signal the operation of this scenario, though we speculate this could be a new channel to make rapidly rotating Be stars in binaries without any prior mass transfer. 

The latter scenario of very slowly rotating stars in close binaries has already been observed in multiple systems. For example, the short-period binary HD~201433 \citep{kallinger:17}, containing an SPB star in a $P=3.3 \, {\rm d}$ orbit with a lower mass companion star, may result from inverse tides. The asteroseismically measured internal rotation rate of this star is $P_{\rm rot} \sim 300 \, {\rm d}$, two orders of magnitude longer than the orbital period. The configuration of this binary is similar to the fiducial binary described in Section \ref{numeric}, demonstrating the feasibility of inverse tides for realistic binary and oscillation mode properties. 

In that system, our models indicate the largest amplitude observed modes in the primary usually have amplitudes larger than the tidally forced amplitudes, depending on the resonant detuning. Hence, for the largest amplitude self-excited modes, the value of $x_{\rm f} < 1$, and inverse tides would not occur. However, the tidal interaction is dominated by the mode with frequency $\omega \approx 2 \Omega$, which likely has a smaller non-linear saturation amplitude than the largest amplitude modes, and hence could have $x_{\rm f} \approx 1$ such that inverse tides occur.

All of our results apply equally well to $\gamma$~Dor stars, and in fact several very slowly rotating $\gamma$-Dor stars in close binaries were recently discovered in \cite{li:20}. These include KIC~4142768 ($P_{\rm orb} = 14.0 \ {\rm d}$, $P_{\rm rot} = 2700^{+1300}_{-660} \, {\rm d}$, see also \citealt{guo:19}), KIC~8197761 ($P_{\rm orb} = 9.87 \, {\rm d}$, $P_{\rm rot} = 1190^{+44}_{-41} \, {\rm d}$ from \citealt{li:20} or $P_{\rm rot} = 301^{+3}_{-3} \, {\rm d}$ from \citealt{sowicka:17}), KIC~8429450 ($P_{\rm orb} = 2.71 \, {\rm d}$, $P_{\rm rot} = 38^{+128}_{-17} \, {\rm d}$), KIC~9850387 ($P_{\rm orb} = 2.75 \, {\rm d}$, $P_{\rm rot} = 190^{+74}_{-42} \, {\rm d}$), and KIC~4480321 ($P_{\rm orb} = 9.17 \, {\rm d}$, $P_{\rm rot} = 142^{+14}_{-14} \, {\rm d}$, see \citealt{lampens:18} and \citealt{li:20b}). We note that KIC~4142768 is a heartbeat star in an eccentric orbit, with a periastron orbital period of $P_{\rm peri} = (1-e)^{3/2} P_{\rm orb} \approx 3.8 \, {\rm d}$, which is more relevant when considering the strength of tidal forces. It is possible that eccentricity pumping via inverse tides can form eccentric heartbeat star systems with self-excited g~modes like KIC~4142768, but we have not studied this possibility here.

Examining the above list of extremely slow core rotators, we see that all of them have periastron orbital periods within a factor of 2 of $P_{\rm peri} = 5 \, {\rm d}$. In contrast, systems with shorter periastron periods usually have cores rotating nearly synchronously with the orbit, while systems with longer periastron periods have core rotation rates similar to single stars \citep{li:20}. This distribution is roughly in line with the prediction of inverse tides. In very tight binaries, the tidally forced amplitudes are larger than the modes' non-linear saturation amplitudes, such that $x_{\rm f} \gg 1$, and tides operate in the usual direction of enforcing spin-orbit synchronization. In wider binaries, the tidally forced amplitudes are smaller such that $x_{\rm f} \ll 1$, but tides are not strong enough to enforce spin-orbit synchronization.

Main sequence stars with $P_{\rm peri} \sim 2-10 \, {\rm d}$ are in the sweet spot where tidally excited modes can have $x_{\rm f} \sim 1$ such that inverse tides can operate, potentially slowing the near core rotation rate to values near zero as discussed in Section \ref{results}. We note that the final rotation periods found in Figure \ref{OrbEvol} are often tens to hundreds of days (depending on the initial misalignment angle), similar to the observed core rotation periods in the binaries discussed above. A prediction of our model is that those stars' spin axes are misaligned with respect to their orbits. However, we caution that our models only include one oscillation mode and do not account for changes in tidal forcing frequency, so more detailed work will be needed to compute robust spin period predictions.

In principle, inverse tides could also operate in exoplanetary systems in short-period orbits around a pulsating host star.
%In that case, the spin and orbital AM can be comparable, and the evolution could proceed in a different fashion.
However, in most cases, we do not expect inverse tides to operate for exoplanet systems because the tidally forced amplitudes are orders of magnitude smaller (due to the smaller planet mass). With the possible exception of very massive planets in very short-period orbits, we thus expect $x_{\rm f} \ll 1$ such that inverse tides do not operate in exoplanet systems.

Our work has assumed rigid stellar rotation, but at least two of the above slowly rotating stars (HD~201433 and KIC~8197761, \citealt{kallinger:17,sowicka:17}) show surface rotation apparently synchronized with the orbit, implying large differential rotation between the slowly rotating core and moderately rotating surface layers. This shear may be sustained by a competition between inverse tides and normal tides operating for different tidally excited modes, along with ongoing AM transport through other processes. Future work should also attempt to account for the differential rotation that naturally develops when any pulsation mode is excited \citep{townsend:18}. We expect nearly rigid rotation to be enforced due to magnetic torques in most of the stellar interior \citep{spruit:02,fuller:19}, but evidently large shear can develop in the outer few percent of the star's radius, perhaps due to the very small mass/inertia of these layers, which may allow them to be easily torqued into synchronous rotation.

Along these lines, \cite{lee:93} showed that even damped oscillation modes can drive rotation opposite to the orbital motion in near-surface mode driving regions. If this process dominated the tidal interaction, we might expect to see that the surface layers of stars in close binaries rotate non-synchronously, while the deep interior rotates synchronously. Observations indicate the opposite. %However, at least for the two slow rotators with surface rotation measurements discussed above, observations indicate the opposite: synchronized surface layers and asynchronous core rotation.
This could be caused by some form of ``normal" tidal friction (e.g., convective viscosity in sub-surface convection zones) that keeps the near-surface layers synchronized, while inverse tides drive asynchronicity in the interior. More surface rotation measurements should be performed for pulsating binaries to test this possibility.

Both \cite{Ogilvie2012} and \citep{duguid:20} computed the effective viscosity due to the interaction of tidal forcing and convective turbulence, finding that \textit{negative} viscosities were possible in some circumstances, potentially producing inverse tidal evolution qualitatively similar to that discussed here. While this is unlikely to explain the systems discussed above (which have radiative interiors apart from very thin convective envelopes and small convective cores), future work should examine whether inverse tides arising from negative convective viscosity can occur in Nature.

\section*{Acknowledgments}

I thank the anonymous referee for a very careful reading and thoughtful report. I am thankful for support through an Innovator Grant from The Rose Hills Foundation, and the Sloan Foundation through grant FG-2018-10515.

\section*{Data Availability}
Data and source code is available upon request to the authors.

\bibliography{CoreRotBib,library}

\end{document}